\documentclass[draftclsnofoot,onecolumn]{IEEEtran}
\ifCLASSINFOpdf

\else

\fi

\usepackage{CJK}
\usepackage{algorithm}
\usepackage{algorithmic}
\usepackage{amsmath}
\usepackage{amssymb}
\usepackage{psfrag}
\usepackage{graphicx}

\usepackage{epstopdf}
\usepackage{multirow}
\usepackage{stfloats}
\usepackage{cite}
\usepackage{subcaption}
\usepackage{color}
\usepackage[normalem]{ulem}
\usepackage{arydshln}
\usepackage{enumerate}
\usepackage{bbm}
\usepackage{hyperref}

\newcommand{\bm}[1]{\mbox{\boldmath{$#1$}}}

\begin{document}

\title{Channel Estimation for Reconfigurable Intelligent Surface-Assisted Cell-Free Communications }
\author{Songjie Yang, Chenfei Xie, Mingwei Wang and Zhongpei Zhang
\thanks{

All authors are with the National Key Laboratory of Science and Technology on Communications, University of Electronic Science and Technology of China (UESTC), Chengdu 611731, China (e-mails:	yangsongjie@std.uestc.edu.cn; 201911220505@std.uestc.edu.cn; 15030786778@163.com; zhangzp@uestc.edu.cn).
}
}
\maketitle
\begin{abstract}
Recent research has focused on reconfigurable intelligent surface (RIS)-assisted cell-free systems with the goal of enhancing coverage and lowering the cost of cell-free networks. However, current research makes the assumption that the perfect channel state information is known. Channel acquisition is, certainly, a difficulty in this case. This work is aimed at investigating RIS-assisted cell-free channel estimation. Toward this end, two unique characteristics are pointed out: 1) For all users, a common channel exists between the base station (BS) and the RIS; and 2) For all BSs, a common channel exists between the RIS and the user. Based on these two characteristics, cascaded and two-timescale channel estimation concerns are studied. Subsequently, two solutions for tackling with the two issues are presented respectively: a three-dimensional multiple measurement vector (3D-MMV)-based compressive sensing technique and a multi-BS cooperative pilot-reduced methodology. Finally, simulations illustrate the effectiveness of the schemes we have presented.

\end{abstract}


\IEEEpeerreviewmaketitle

\section{Introduction}
Cell-free network techniques have been suggested as a user-centric paradigm for coordinating numerous randomly distributed base stations (BSs) without regard to cell borders in order to serve all network users concurrently. Benefiting from the distributed deployment of BSs, cell-free networks may boost diversity and coverage. Until date, cell-free networks have attracted considerable interest in academia, with a particular emphasis on downlink precoding \cite{DP1}, power control \cite{PC1} and non-orthogonal multiple-access (NOMA)\cite{NOMA1,NOMA2}.

 However, the conventional cell-free system needs a large-scale deployment of BSs, which results in an unsatisfactory energy efficiency performance owing to the high costs of both hardware and power sources.
To address this problem, the reconfigurable intelligent surface (RIS) technique has evolved as a potentially cost-effective methodology that works by establishing favorable propagation circumstances between the BS and users. More precisely, the RIS is a reflective array comprised of numerous low-cost and energy-efficient passive reflective components capable of implementing passive beamforming with changeable parameters.
 In recent years, a variety of published research has focused on using the RIS as a reflector to improve wireless communications, such as optimizing the spectral efficiency \cite{RIS2,RIS3}.
\begin{figure}
	\centering
	\includegraphics[width = 0.4\textwidth]{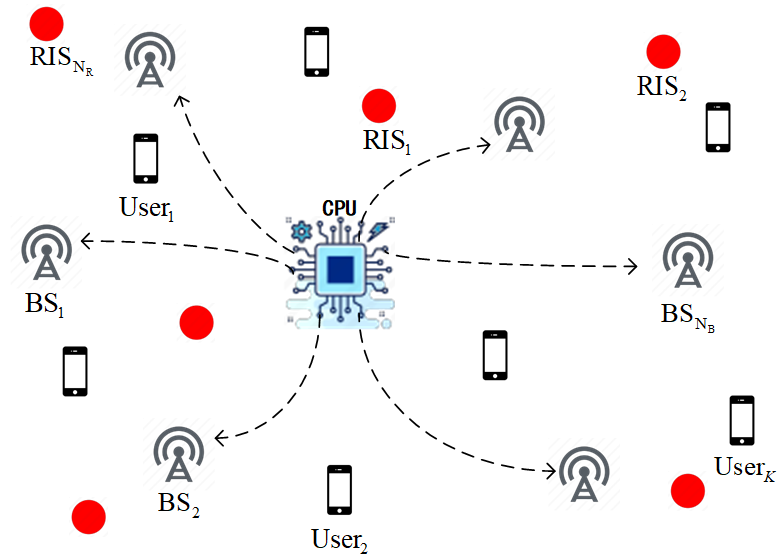}
	\caption{A RIS-assisted cell-free communication system.}
	\label{Fig1SystemModel}
\end{figure}

So far, several studies on RIS-assisted cell-free networks have been presented, with a focus on capacity \cite{capacity}, energy efficiency \cite{EE2,EE3}, joint precoding \cite{precoding} and simultaneous wireless information and power transfer (SWIPT) \cite{SWIPT}. However, the above studies assume that perfect channel state information (CSI) is known. Indeed, channel estimation with RISs is challenging since the RIS is a passive device with numerous passive reflective elements that cannot actively transmit/receive or process signals.

Despite all this, a variety of studies in the scenario of single-BS single-RIS channel estimation have been documented in \cite{binary1,binary2,PARAFAC1,PARAFAC2,CS1,CS2,CS3,CS4,two-timescale}, which can be broadly classified into cascaded and two-timescale channel estimation. For cascaded channel estimation, binary reflection, parallel factor (PARAFAC) decomposition, and compressive sensing (CS) strategies are widely investigated. Specifically, the central concept of binary reflection methods is to utilize an ON/OFF protocol for RIS elements \cite{binary1,binary2}. This ON/OFF switching mechanism is expensive as it needs a separate amplitude control for each reflecting element. Inspired by PARAFAC’s promising results in channel estimation for relay systems, PARAFAC-based channel estimation approaches for RISs were presented in \cite{PARAFAC1,PARAFAC2}, where the individual channel was refined iteratively using bilinear alternating least squares.
By exploiting the characteristics of the Khatri-Rao product, the Kronecker product, and the sparsity of wireless channels, the authors in \cite{CS1} constructed cascaded channel estimation as a CS issue. Furthermore, CS-based cascaded channel estimation schemes emerged in \cite{CS2,CS3,CS4}. 
\begin{figure}
	\centering
	\includegraphics[width = 0.45\textwidth]{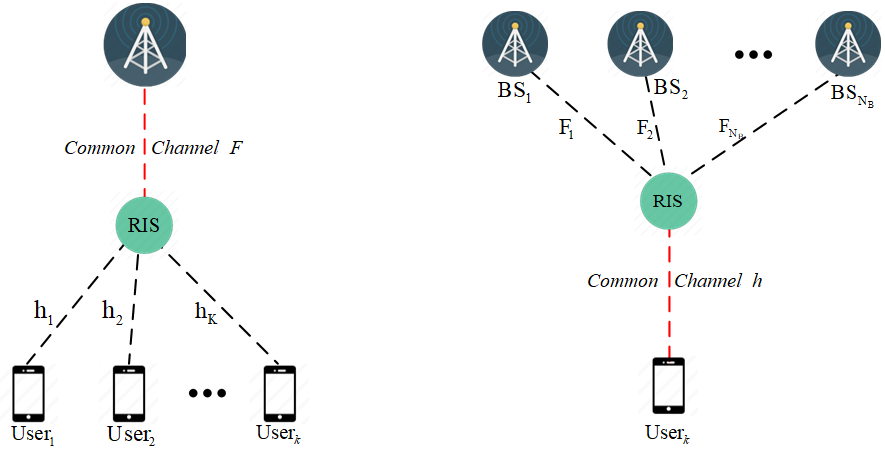}
	\caption{Multi-user and multi-BS channels characteristics.}
	\label{two_characteristics}
\end{figure}
In the spirit of using the slow time-varying property of BS-RIS channels (since the positions of the BS and the RIS are fixed, the channel between them is slow time-varying) and the fast time-varying property of RIS-user channels, a two-timescale channel estimation scheme was suggested in \cite{two-timescale}, where the first and second timescale were used to estimate the slow and fast time-varying channels, respectively. Additionally, two-timescale channel estimation requires fewer pilots than cascaded channel estimation. However, the research to date has not investigated channel estimation with RIS-assisted cell-free systems (i.e., multi-BS multi-RIS scenarios). This motivates us to suggest RIS-assisted cell-free channel estimation strategies.

Taking the preceding into consideration, the purpose of this research is to develop a framework for RIS-assisted cell-free channel estimation from two perspectives: cascaded and two-timescale channel estimation. Toward this end, we'd like to point out two unique characteristics of the multi-BS multi-user scenario as follows:
\begin{itemize}
\item {\bf{Characteristic 1}:} Given the BS and the RIS, multi-user channels share a common part (BS-RIS), as seen on the left-hand side of \figurename{\ref{two_characteristics}}.

\item {\bf{Characteristic 2}:} Given the user and the RIS, multi-BS channels share a common part (RIS-user), as seen on the right-hand side of \figurename{\ref{two_characteristics}}.
\end{itemize}

These two characteristics imply that joint multi-user/BS signal processing is capable of delivering efficient estimation methods\footnote{CS-based techniques are the subject of this investigation. Indeed, other technologies can develop efficient alternatives using these two characteristics.}. In this manner, we present a three-dimensional multiple measurement vector (3D-MMV)-based CS framework for cascaded channel estimation, followed by the development of a 3D-MMV look ahead orthogonal match pursuit (3D-MLAOMP) algorithm based on tensor contraction.
Besides, we propose a pilot-reduced two-timescale channel estimation strategy via multi-BS cooperation.

{\emph {Notations}}:
 ${\left(  \cdot  \right)}^{ *}$, ${\left(  \cdot  \right)}^{ T}$ and ${\left(  \cdot  \right)}^{ H}$ denote conjugate, transpose, conjugate transpose, respectively. ${{\mathbf{A}}^\dag }$ is the Moore-Penrose pseudoinverse matrix. $\Vert\mathbf{a}\Vert_0$, $\Vert\mathbf{a}\Vert_2$, $\Vert\mathbf{A}\Vert_{\ell_1}$ and $\Vert\mathbf{A}\Vert_F$ denote $\ell_0$, $\ell_2$, $\ell_1$ and Frobenius norm, respectively. $\vert \Lambda \vert$ is the cardinality of set $\Lambda$. $\otimes$ is the Kronecker product, $\left\langle \cdot,\cdot\right\rangle_F$ is the Frobenius inner product and $\left\langle \cdot |\cdot\right\rangle$ represents tensor contraction. $[\mathbf{a}]_{i}$, $[\mathbf{A}]_{i,j}$ and $[\mathcal{A}]_{i,j,k}$ denote the $i$-th element of vector $\mathbf{a}$, the $(i,j)$-th element of matrix $\mathbf{A}$ and the $(i,j,k)$-th element of tensor $\mathcal{A}$, respectively. Finally, $\mathbf{I}_M$ denotes the $M$-by-$M$ identity matrix.

%

\section{System and Channel Model}\label{sec:System Model}
\subsection{Scenario Caption and Signal Model}

As shown in Fig.~\ref{Fig1SystemModel}, we consider a time division duplex (TDD) massive MIMO system, where $N_{B}$ BSs with the assistance of $N_{R}$ RISs serve $K$ users. The $m$-th BS and the $n$-th RIS are equipped with $J_m$ antennas and $L_n$ reflective elements, respectively, where $m=1,2,\cdots,N_{B}$ and $n=1,2,\cdots,N_{R}$. The users are all equipped with a single antenna. In such a scenario, RISs operate as antenna arrays located distant from the BS to increase capacity and provide larger coverage at a low cost, hence enhancing cell-free communications. Additionally, a central processing unit (CPU) is arranged to facilitate the joint signal processing of multiple BSs. All RISs are controlled by BSs or the CPU by wired control.

We consider quasi-static block-fading channels, the channel matrices from the $m$-th BS to the $n$-th RIS and from the $n$-th RIS to the $k$-th user are represented by $\mathbf{F}_{mn}\in\mathbb{C}^{L_n\times J_m}$ and $\mathbf{h}_{nk}^H\in\mathbb{C}^{1\times L_n}$, respectively. For the $n$-th RIS, the reflection coefficient matrix is expressed as $\mathbf{V}_n={\rm diag}(\mathbf{v}_n)\in\mathbb{C}^{L_n\times L_n}$ with $\mathbf{v}_n=[v_{n,1},v_{n,2},\cdots,v_{n,L_n}]^T$, where $v_{n,l}=\rho_{n,l}e^{j\theta_{n,l}}$ is the reflection coefficient of the $l$-th element of the $n$-th RIS, $\rho_{n,l}=1$ and $\theta_{n,l}$ are amplitude and phase of it, respectively. Then the channel matrix from the $m$-th BS to the $k$-th uer caused by the $n$-th RIS is $\mathbf{h}^H_{nk}\mathbf{V}_n\mathbf{F}_{mn}\in\mathbb{C}^{1\times J_m}$. For the sake of clarity, it is assumed that each RIS has the same number of reflection elements, i.e., $L_1=L_2=\cdots=L_{N_{R}}=L$.
Owing to the property of channel reciprocity in TDD systems, the CSI of the downlink can be attained by uplink channel estimation.
For uplink channel estimation, the key is to design the orthognal pilot sequences and reflection coefficients to measure the channel. Assume that the uplink channel estimation stage cosists of $Q $ sub-frames for RISs and each sub-frame accommodates $T$ symbol durations, where $T\geq K$.  

 The reflection vector of the $n$-th RIS in the $q$-th sub-frame is denoted by $\mathbf{v}^q_n$, the pilot sequence of the $k$-th user is $\mathbf{s}^H_k=[s_{k,1},s_{k,2},\cdots,s_{k,T}]\in\mathbb{C}^{1\times T}$ and satisfies $\mathbf{s}^H_k\mathbf{s}_j=0, \ k\neq j$ and $\mathbf{s}^H_k\mathbf{s}_k=\sigma^2_p T$, where $\sigma_p^2$ is the transmit power of each user. Since the direct channel between the BS and the user can be estimated simply by conventional techniques via the setting that all RISs are turned off, the direct channel estimation is not considered. In the $q$-th sub-frame, the received signal ${{\mathbf{Y}}_{m,q}}\in\mathbb{C}^{J_m\times T}$ at the $m$-th BS  is written as
\begin{equation}
	{{\mathbf{Y}}_{m,q}} = \sum\limits_{k = 1}^K \sum\limits_{n = 1}^{N_{R}} {{{\mathbf{F}}_{mn}^{H}}\mathbf{V}^q_n{{\mathbf{h}}_{nk}}{{\bm s}_{k}^{H}}}  + {{\bm W}_{m,q}},
\end{equation}
where ${{\bm W}_{m,q}}\in\mathbb{C}^{J_m\times T}$ is the received Gaussian noise following $\mathcal{CN}(0,\sigma_n^2\mathbf{I}_{J_m})$.
Since the orthogonal pilot sequences are employed\footnote{This work considers low mobility scenarios with orthogonal pilot sequences. Indeed, our proposed framework is also applicable to not-orthogonal pilot sequences.}, for the $k$-th user, we have
\begin{equation}
	  \tilde{\mathbf{y}}_{m,q,k}=\frac{1}{\sigma_p^2 T}\mathbf{Y}_{m,q}\mathbf{s}_k 
	  =  \sum\limits_{n = 1}^{N_{R}} {{{\mathbf{F}}_{mn}^{H}}\mathbf{V}^q_n{{\mathbf{h}}_{nk}}}+\widetilde{\mathbf{W}}_{m,q},
\end{equation}
where $\widetilde{\mathbf{W}}_{m,q}=\frac{1}{PT}{{\bm W}_{m,q}}$. By collecting the signals of $Q$ sub-frames via raw stacking, the received signal $\widetilde{\mathbf{Y}}_{m,k}\in\mathbb{C}^{J_mQ\times1}$ is expressed as
\begin{equation}\label{Y}
		\widetilde{\mathbf{Y}}_{m,k}
		= \begin{bmatrix}\sum\limits_{n = 1}^{N_{R}} {{{\mathbf{F}}_{mn}^{H}}\mathbf{V}^1_n{{\mathbf{h}}_{nk}}} \\
			\sum\limits_{n = 1}^{N_{R}} {{{\mathbf{F}}_{mn}^{H}}\mathbf{V}^2_n{{\mathbf{h}}_{nk}}}\\
			\vdots\\
			\sum\limits_{n = 1}^{N_{R}} {{{\mathbf{F}}_{mn}^{H}}\mathbf{V}^Q_n{{\mathbf{h}}_{nk}}}
		\end{bmatrix} +
		\begin{bmatrix}\widetilde{\mathbf{W}}_{m,1}\\
					\widetilde{\mathbf{W}}_{m,2}\\
					\vdots\\
					\widetilde{\mathbf{W}}_{m,Q}
		\end{bmatrix}.
\end{equation}

\subsection{Channel Model}
Postulate that all BSs and RISs are equipped with a uniform linear array (ULA). The widely adopted physical channels \cite{CS1,CS2} characterizing the geometrical structure are expressed as
\begin{equation}\label{eqhF}
\begin{aligned}
	{\mathbf{F}_{mn}}\!= \!\sqrt {\frac{{L_n\!J_m}}{{P_{f,m,n}}}} \!\sum\limits_{p = 1}^{{P_{f,m,n}}}\! &{\beta _{m,n,p}}{\mathbf{a}}_L\left(  \frac{2\pi d}{\lambda}\sin(\vartheta_{m,n,p})\right)\\&
	\times{{\mathbf{a}}_D^{H}}	
	\left(  {\frac{2\pi d}{\lambda}\sin(\phi_{m,n,p})}  \!\right),
\end{aligned}
\end{equation}
\begin{equation}
	{{\mathbf{h}}_{nk}}\!=\!\sqrt {\frac{{L}_n}{{P_{h,n,k}} }}\sum\limits_{b = 1}^{{P_{h,n,k}}} {{\gamma_{n,k,b}}{\mathbf{a}}_L\left(  \frac{2\pi d}{\lambda}{\sin(\varphi_{n,k,b})} \right)},\label{eqhk}
\end{equation}
where $P_{f,m,n}$, $P_{h,n,k}$ are the number of paths of  ${\mathbf{F}_{mn}}$ and ${{\mathbf{h}}_{nk}}$, respectively, $\beta_{m,n,p}$ and $\gamma_{n,k,b}$ are the complex gains of the $p$-th and $b$-th path of the two channels, respectively. Moreover, $\phi_{m,n,p}$ and $\varphi_{n,k,b}$ represent AoDs from the $m$-th BS to the $n$-th RIS and from the $n$-th RIS to the $k$-th user, respectively. $\vartheta _{m,n,p}$ is the AoA from the $m$-th BS to the $n$-th RIS. In addition, $d$ is the antenna inter-element spacing, and $\lambda$ denotes the antenna wavelength. $\mathbf{a}_D(\cdot)$ and $\mathbf{a}_L(\cdot)$ are ULA steering vectors of BSs and RISs, respectively. Ignoring the subscript without losing generality, we have
\begin{equation}\label{linear}
	\mathbf{a}(\theta)=	\sqrt{\frac{1}{N}}[1,e^{j\theta},\cdots,e^{j (N-1)\theta}]^{ T},
\end{equation}
where $N$ is the number of antenna elements.

\section{Cascaded Channel Estimation}
One straightforward approach to cascaded channel estimation is the least squares (LS) algorithm employing much pilots for adjusting reflecting matrices.
Denote the cascaded channel among the $m$-th BS, the $k$-th user and the $n$-th RIS as $\mathbf{G}_{mkn}={\rm diag}(\mathbf{h}_{nk}^H)\mathbf{F}_{mn}\in\mathbb{C}^{L\times J_m}$, and  $\overline{\mathbf{Y}}_{m,k}\in\mathbb{C}^{J_m\times Q}$ as the colmun stacking form of the collected signals. Then Eqn. (\ref{Y}) can be rewritten as
\begin{equation}
	\begin{aligned}
	\overline{\mathbf{Y}}_{m,k}=&\sum\limits_{n = 1}^{N_{R}} \mathbf{G}_{mkn}^H\overline{\mathbf{V}}_n+\overline{\mathbf{W}}_{m,q} \\
	=&\begin{bmatrix}
	\mathbf{G}_{mk1}^H, \cdots, \mathbf{G}_{mkN_{R}}^H
	\end{bmatrix} \begin{bmatrix}\overline{\mathbf{V}}_1\\
	\vdots\\
		\overline{\mathbf{V}}_{N_{R}}
	\end{bmatrix} +\overline{\mathbf{W}}_{m,q},
	\end{aligned}
\end{equation}
where $\overline{\mathbf{V}}_n=[{{\mathbf{v}}^1_{n}},{{\mathbf{v}}^2_{n}},\cdots,{{\mathbf{v}}^Q_{n}}]\in\mathbb{C}^{L\times Q}$, $\overline{\mathbf{W}}_{m,q}=[\widetilde{\mathbf{W}}_{m,1},\widetilde{\mathbf{W}}_{m,2},\cdots,\widetilde{\mathbf{W}}_{m,Q}]\in\mathbb{C}^{J_m\times Q}$.

According to the LS channel estimation method, the estimated cascaded channels $\{\widehat{\mathbf{G}}_{mkn}\}_{n=1}^{N_R}$ are give by
\begin{equation}
	\begin{aligned}&
	\begin{bmatrix}
		\widehat{\mathbf{G}}_{mk1}^H,\cdots, \widehat{\mathbf{G}}_{mkN_{R}}^H
	\end{bmatrix}\\
=&\overline{\mathbf{Y}}_{m,k}\begin{bmatrix}
		\overline{\mathbf{V}}_1\\
		\vdots\\
		\overline{\mathbf{V}}_{N_{R}}
	\end{bmatrix}^H \begin{pmatrix}
\begin{bmatrix}
	\overline{\mathbf{V}}_1\\
	\vdots\\
	\overline{\mathbf{V}}_{N_{R}}
\end{bmatrix}\begin{bmatrix}
\overline{\mathbf{V}}_1\\
\vdots\\
\overline{\mathbf{V}}_{N_{R}}
\end{bmatrix}^H
\end{pmatrix}^{-1}\\
=&\overline{\mathbf{Y}}_{m,k}\begin{bmatrix}
	\overline{\mathbf{V}}_1\\
	\vdots\\
	\overline{\mathbf{V}}_{N_{R}}
\end{bmatrix}^H \begin{bmatrix}
		\overline{\mathbf{V}}_1\overline{\mathbf{V}}_1^H &\cdots & \overline{\mathbf{V}}_1\overline{\mathbf{V}}_{N_R}^H\\
		\vdots &\ddots&\vdots\\
		\overline{\mathbf{V}}_{N_R}\overline{\mathbf{V}}_1^H&
		\cdots&
		\overline{\mathbf{V}}_{N_R}\overline{\mathbf{V}}_{N_R}^H
\end{bmatrix}^{-1}.
\end{aligned}
\end{equation}

 It can be observed that the case of $\overline{\mathbf{V}}_i\overline{\mathbf{V}}_j^H=\mathbf{O}, \ i\neq j$, where $\mathbf{O}$ represents null matrix, is the time switching (TS) mode, which lies on switch to control each RIS to work in turn. This way may result in low energy consumption since only a RIS is turned on for channel estimation at a time.
 Notably, the training overhead is too high for LS estimation, i.e., $Q\geq N_RL$ is required. Thus, CS-based channel estimation with low training overhead is more preferred, especially the size of RISs is large. Following that, we'll explore efficient channel estimation in the TS mode.

In the CS-based channel estimation, the key is the virtual channel representation based on the discrete Fourier transformation (DFT) dictionary. The cascaded channel $\mathbf{G}_{mkn}$ is written as
\begin{equation}\label{GVCR}
	\begin{aligned}
\mathbf{G}_{mkn}=&\sum_{p=1}^{P_{f,m,n}}\sum_{b=1}^{P_{h,n,k}}\sqrt{\frac{L_n^2J_m}{P_{f,m,n}P_{h,n,k}}}\beta_{m,n,p}\gamma_{n,k,b}\times\\
&{\mathbf{a}}_L({\vartheta}_{m,n,p}-{\varphi}_{n,k,b}){\mathbf{a}}^H_D({\phi}_{m,n,p}),
	\end{aligned}
\end{equation} 
where $({\vartheta}_{m,n,p}-{\varphi}_{n,k,b})$ can be regard as user's cascaded AoA.

 A virtual compact form of Eqn. (\ref{GVCR}) is
\begin{equation}\label{CGVCR}
	\mathbf{G}_{mkn}=\mathbf{A}_{\rm R}\mathbf{X}_{mkn}\mathbf{A}_{{\rm T},m}^H,
\end{equation}
where $\mathbf{A}_{\rm R}$ and $\mathbf{A}_{\rm T}$ represent the over-complete DFT dictionary of the user's cascaded AoA and the BS's AoD, respectively, and can be written as
\begin{equation}\label{D_A}
	\mathbf{A}_{\rm R}=[{\mathbf{a}}_{L}(-1),{\mathbf{a}}_{L}(-1+\frac{2}{G_r}),\cdots,{\mathbf{a}}_{L}(1-\frac{2}{G_r})]\in\mathbb{C}^{L\times G_r},
\end{equation}
\begin{equation}
	\mathbf{A}_{{\rm T}, m}=[{\mathbf{a}}_{D}(-1),{\mathbf{a}}_{D}(-1+\frac{2}{G_t}),\cdots,{\mathbf{a}}_{D}(1-\frac{2}{G_t})]\in\mathbb{C}^{J_m\times G_t},
\end{equation}
where $G_r$ and $G_t$ are the number of the AoAs and AoDs on the grid. Additionally, $\mathbf{X}_{mkn}$ represents the sparse matrix accommodating entries of AoAs, AoDs and gains.

For clarity, we ignore the subscript of the RIS due to the TS mode is adopted, and denote $\overline{\mathbf{V}}\in\mathbb{C}^{L\times \bar{Q}}$ by the current RIS matrix, where $\bar{Q}$ is the number of sub-frames for estimating the cascaded channel of the current RIS, then the CS-based channel estimation issue is formulated as
\begin{equation}\label{CS1}
	\begin{aligned}
	&\underset{\mathbf{X}_{mk}}{\rm arg \ min} \ \Vert \mathbf{X}_{mk} \Vert_0 \\
{\rm s.t.} \ \Vert\overline{\mathbf{Y}}^H_{m,k}&-\overline{\mathbf{V}}^H\mathbf{A}_{\rm R}\mathbf{X}_{mk}\mathbf{A}_{{\rm T},m}^H\Vert^2_F<\epsilon,
	\end{aligned}
\end{equation}
where $\epsilon$ is the recovery precision.

Unlike the one-dimensional (1D) CS approach used in \cite{CS1}, whereas ${\rm vec}(\mathbf{X}_{mk})$ was recovered without exploiting the two-dimensional (2D) sparsity of $\mathbf{X}_{mk}$, resulting in high computational complexity, the authors in \cite{CS2} employed the MMV-based method to recover $\widetilde{\mathbf{X}}_{mk}=\mathbf{X}_{mk}\mathbf{A}_{{\rm T},m}^H$ with row-group sparsity\footnote{In a matrix with row-group sparsity, there are only a few rows with non-zero elements, and all of the rest of the rows are zero.}, and then attained $\widehat{\mathbf{G}}_{mk}=\mathbf{A}_{\rm R}\widetilde{\mathbf{X}}_{mk}$. 

\begin{figure}
	\centering
	\includegraphics[width = 0.339\textwidth]{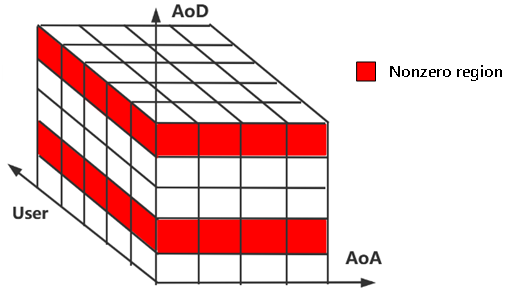}
	\caption{Diagrammatic sketch of the 3D-MMV structure.}
	\label{3DMMV}
\end{figure}
In accordance with {\bf{Characteristic 1}}, the compressive channel estimation of each user can be divided into common part (AoDs) and individual part (AoAs). This specifies that all users can collaborate to estimate AoDs, and then individually estimate AoAs. To begin, we frame the AoD recovery issue as a 3D-MMV problem shown in \figurename{\ref{3DMMV}}, where AoDs are spread as matrix slices in the red region.
 By re-organizing the received signals $\{\overline{\mathbf{Y}}_{m,k}\}_{k=1}^K$ into a three-order tensor $\mathcal{Y}^m\in\mathbb{C}^{J_m\times \bar{Q}\times K}$ and regarding the dictionary $\mathbf{A}_{\rm T}$ as a two-order tensor, we utilize the tensor contraction operator \cite{contraction} to formulate the 3D-MMV problem:
\begin{equation}
	\begin{aligned}
	\underset{\mathcal{Z}^{m}}{\rm arg \ min} \ & \sum_{i=1}^{G_t}\Vert \left\langle [\mathcal{Z}^{m}]_{i,:,:},[\mathcal{Z}^{m}]_{i,:,:}^H \right\rangle_F \Vert_0 \\
	{\rm s.t.} \ &\Vert \mathcal{Y}^{m}-\left\langle\mathbf{A}_{\rm T}|\mathcal{Z}^m\right\rangle\Vert^2_F<\epsilon,
	\end{aligned}
\end{equation}
where $\mathcal{Z}^m\in\mathbb{C}^{G_t\times \bar{Q}\times K}$ and $[\mathcal{Z}^m]_{:,:,k}$ consists of the $k$-th user's estimated signal whose nonzero rows correspond to the AoD indexes in the dictionary $\mathbf{A}_{\rm T}$. Additionally, the tensor contraction between $\mathbf{A}_{\rm T}$ and $\mathcal{Z}^m$ is expressed as $\left\langle\mathbf{A}_{\rm T}|\mathcal{Z}^m\right\rangle=\sum_{g=1}^{G_t}[\mathbf{A}_{\rm T}]_{j,g}[\mathcal{Z}^m]_{g,q,k}$.

Upon $[\mathcal{Z}^m]_{:,:,k}\approx\mathbf{X}_{mk}^H\mathbf{A}^H_{\rm R}\overline{\mathbf{V}}$ is recovered for all BSs, the nonzero row indexes are recorded in set $\Omega^{AoD}_m$. Then we attain the low-dimensional signal $[\widehat{\mathcal{Z}}^m]_{:,:,k}\in\mathbb{C}^{P_{AoD}\times Q}$ by removing all zero rows of $[\mathcal{Z}^m]_{:,:,k}$, where $P_{AoD}$ represents the number of nonzero rows of the recovered signal. Up to this point, each user's AoAs can be estimated by 1D CS methods. For concreteness, we first vectorize $[\widehat{\mathcal{Z}}^m]_{:,:,k}$, then use the dictionary $\overline{\mathbf{V}}^T\mathbf{A}_{\rm R}^*\otimes\mathbf{I}_{P_{AoD}}$ to recover ${\mathbf{x}}_{mk}$, where each nonzero entry corresponds the AoA and path gain. That is 
\begin{equation}\label{AOA}
		\begin{aligned}
		&\underset{\mathbf{x}_{mk}}{\rm arg \ min} \ \Vert \mathbf{x}_{mk} \Vert_0 \\
		{\rm s.t. } \ &\Vert \hat{\mathbf{z}}_{mk}-\widetilde{\mathbf{A}}_{\rm R}\mathbf{x}_{mk}\Vert^2_F<\epsilon,
	\end{aligned}
\end{equation}
where $\hat{\mathbf{z}}_{mk}={\rm vec}([\widehat{\mathcal{Z}}^m]_{:,:,k})$ and $\widetilde{\mathbf{A}}_{\rm R}=\overline{\mathbf{V}}^T\mathbf{A}_{\rm R}^*\otimes\mathbf{I}_{P_{AoD}}$.
By denoting $\Omega_{mk}^{AoA}$ and $\Omega^{Ga}_{mk}$ as the AoA and gain sets, the cascaded channel $\mathbf{G}_{mk}$ can be reconstructed via $\Omega_{mk}^{AoA}$, $\Omega_{m}^{AoD}$ and $\Omega^{Ga}_{mk}$.

 \begin{algorithm}[!t] 
	\caption{3D-MLAOMP for Cascaded Channel Estimation} 
	\label{3dce}      
	\begin{algorithmic}[1] 
		\footnotesize{
			\REQUIRE {Received signals $\mathcal{Y}^m$, reflecting matrix $\overline{\mathbf{V}}$, dictionaries $\mathbf{A}_{\rm T}$ and $\mathbf{A}_{\rm R}$, and look ahead parameter $U$. }

			\ENSURE {Reconstruction of $\widehat{\mathbf{G}}_{mk}$.} 
			\STATE{\%\% \emph{AoD estimation for all users}}
			\STATE{$\textbf{Initialize:}$  $p=0$, $\Lambda_0=\emptyset$ and $\mathcal{R}^m_0=\mathcal{Y}^m$. 
			}					
			\REPEAT
			\STATE{
				$	p\leftarrow p+1$}.
			\STATE{ \emph{Match Atoms:}}
			\STATE{\ \ \ $\mathcal{I}^m=\left\langle\mathbf{A}_{\rm T}^H|\mathcal{R}^m_{p-1}\right\rangle$, 
				$[\mathbf{g}_t]_g=\Vert[\mathcal{I}^m]_{g,:,:}\Vert_{\ell_1}^2, \ g=1,2,\cdots,G_t$;}  
			\STATE{\ \ \ 
				$\mathbf{u}\leftarrow$ indices of the $U$ largest values of $\mathbf{g}_t$ such that $\mathbf{u}\notin\Lambda_{p-1}$;}
			
			\FOR{$i=1,2,\cdots,U$}
			\STATE{$\mathcal{T}^m=$ {\textbf{Look Ahead Residual}} $(\mathcal{Y}^m,\mathbf{A}_{\rm T},\Lambda_{p-1},[\mathbf{u}]_i)$;}
			\STATE{ $[\mathbf{r}]_i=\Vert\mathcal{T}^m\Vert_F^2$;}
			\ENDFOR
			\STATE{$i_p\leftarrow$ index of the lowest component of $\mathbf{r}$;}
				 \STATE{$\Lambda_p=\Lambda_{p-1}\cup[\mathbf{u}]_{i_p}$;}

			\STATE{\emph{Comupute LS:} $\mathcal{Z}^m_{\Lambda_p}=\left\langle\mathbf{A}^{\dagger}_{{\rm T},\Lambda_p}|\mathcal{Y}^m\right\rangle$;
			}
			
			
			\STATE{
				\emph{Renew Residual:}
				$\mathcal{R}^m_p=\mathcal{Y}^m-\left\langle\mathbf{A}_{{\rm T},\Lambda_p}|\mathcal{Z}^m_{\Lambda_p}\right\rangle$.
				
			}

			\UNTIL{ $\Vert \mathcal{R}_p^m\Vert^2_F<\varepsilon$.}
			\STATE{$\Omega^{AoD}_m\leftarrow\Lambda_{p}$,  $\mathcal{Z}^m\leftarrow\mathcal{Z}^m_{\Lambda_{p}}$, ${P_{AoD}}=p$.}
		}
		\STATE{\%\% \emph{AoA and path gain estimation for each user}.}
		\STATE{$[\widehat{\mathcal{Z}}^m]_{:,:,k}\leftarrow$ removing all zero rows of $[{\mathcal{Z}}^m]_{:,:,k}$;}
		\STATE{$\hat{z}_{mk}={\rm vec}([\widehat{\mathcal{Z}}^m]_{:,:,k})$, $\widetilde{\mathbf{A}}_{\rm R}=\overline{\mathbf{V}}^T\mathbf{A}_{\rm R}^*\otimes\mathbf{I}_{P_{AoD}}$;}
		\STATE{Attain $\Omega^{AoA}_{mk}$ and $\Omega^{Ga}_{mk}$ by solving Eqn. (\ref{AOA}) with the conventional LAOMP algorithm.}
		\STATE{Reconstruct $\widehat{\mathbf{G}}_{mk}$ in light of Eqn. (\ref{CGVCR})}.
		
		\rule[-1pt]{8.25cm}{0.09em}
		\STATE{\textbf{Look Ahead Residual} $(\mathcal{Y},\mathbf{A} ,\Lambda,u)$}
		\STATE{$\textbf{Initialize:}$ $p^\prime=|\Lambda|$, $\Lambda_{p^\prime}=\Lambda\cup u$, $\mathcal{Z}_{\Lambda_{p^\prime}}=\left\langle\mathbf{A}^{\dagger}_{\Lambda_{p^\prime}}|\mathcal{Y}\right\rangle$ and $\mathcal{R}_{p^\prime}=\mathcal{Y}-\left\langle\mathbf{A}_{\Lambda_{p^\prime}}|\mathcal{Z}_{\Lambda_{p^\prime}}\right\rangle$.}
		\REPEAT
		\STATE{$p^\prime\leftarrow p^\prime+1;$}
		\STATE{$[\mathbf{g}_t^\prime]_{g^\prime}=\Vert[\left\langle \mathbf{A}^H|\mathcal{R}_{p^\prime -1}\right\rangle]_{g^\prime,:,:}\Vert_{\ell_1}^2, \ g^\prime=1,2,\cdots,G_t$;}
		\STATE{$u^\prime\leftarrow$ index of the largest value of $\mathbf{g}_t^\prime$ such that $u^\prime\notin 
			\Lambda_{p^\prime-1}$; }
		\STATE{$\Lambda_{p^\prime}=\Lambda_{p^\prime-1}\cup u^\prime$;}
		\STATE{$\mathcal{Z}_{\Lambda_{p^\prime}}=\left\langle\mathbf{A}^{\dagger}_{\Lambda_{p^\prime}}|\mathcal{Y}\right\rangle$, $\mathcal{R}_{p^\prime}=\mathcal{Y}-\left\langle\mathbf{A}_{\Lambda_{p^\prime}}|\mathcal{Z}_{\Lambda_{p^\prime}}\right\rangle$.}
		\UNTIL{$\Vert \mathcal{R}_{p^\prime}\Vert^2_F<\epsilon$}
		\STATE{ \textbf{return} $\mathcal{T}=\mathcal{R}_{\Lambda_{p^\prime}}$}
	\end{algorithmic}
\end{algorithm}
To address the cascaded channel estimation problem utilizing our proposed framework, we adopt an outstanding orthogonal match pursuit variant approach, look ahead orthogonal match pursuit (LAOMP) \cite{LAOMP}, as a starting point, and extend it to the 3D-MLAOMP algorithm, in order to develop Algorithm \ref{3dce}. The core principle of this algorithm is that the index used in the present iteration is chosen based on its future influence on reducing the residual. Specifically, there are two critical steps:
\subsubsection{Match Atoms} As with the MMV issue, we must take various measurements into account while selecting atoms. We will use $\ell_1$ norm as the selection criterion in this case. First of all, the projection of $\mathcal{Y}^m$ onto $\mathbf{A}_{\rm T}$ is attained by tensor contraction:
\begin{equation}
	[\mathcal{I}^m]_{g,q,k}=\left\langle\mathbf{A}_{\rm T}^H|\mathcal{Y}^m\right\rangle=\sum_{j=1}^{J_m}[\mathbf{A}^H_{\rm T}]_{g,j}[\mathcal{Y}^m]_{j,q,k}.
\end{equation}
Then the best atom can be selected by finding the index that maximizes $\mathbf{g}_t$, where $[\mathbf{g}_t]_g=\Vert[\mathcal{I}^m]_{g,:,:}\Vert_{\ell_1}^2, \ g=1,2,\cdots,G_t$.
\subsubsection{Renew Residual} When an atom is determined, the residual must be adjusted in preparation for the next atom selection. Postulate that the present atom support is $\Lambda_p$, where $p$ is present iteration, the LS can be calculated as 
\begin{equation}
	[\mathcal{Z}^m_{\Lambda_p}]_{g,q,k}=\left\langle\mathbf{A}^{\dagger}_{{\rm T},\Lambda_p}|\mathcal{Y}^m\right\rangle=\sum_{j=1}^{J_m}[\mathbf{A}^{\dagger}_{{\rm T},\Lambda_p}]_{g,j}[\mathcal{Y}^m]_{j,q,k},
\end{equation}
where $\mathbf{A}^{\dagger}_{{\rm T},\Lambda_p}=(\mathbf{A}^H_{{\rm T},\Lambda_p}\mathbf{A}_{{\rm T},\Lambda_p})^{-1}\mathbf{A}^H_{{\rm T},\Lambda_p}$ and $\mathbf{A}_{{\rm T},\Lambda_p}$ represents the atoms that correspond to the atom support $\Lambda_p$.
 Following that, the residual $\mathcal{R}^m_p\in\mathbb{C}^{J_m\times\bar{Q}\times K}$ is updated by
\begin{equation}
\mathcal{R}^m_p=\mathcal{Y}^m-\left\langle\mathbf{A}_{{\rm T},\Lambda_p}|\mathcal{Z}^m_{\Lambda_p}\right\rangle.
\end{equation}

Note that {\bf{Characteristic 2}} is related to the AoD from the RIS to users (i.e., $\varphi$), however, we can see that $\varphi$ and $\vartheta$ are coupled in Eqn. (\ref{GVCR}). This precludes the rational use of this characteristic, as a result, we assign each BS performs multi-user cascaded channel estimation independently. For the joint estimation of multiple BSs based on {\bf{Characteristic 2}}, we investigate two-timescale channel estimation with RIS-assisted cell-free systems as follows.

\section{Two-Timescale Channel Estimation}

The central concept of two-timescale channel estimation is that the channel between the BS and the RIS, as well as the channel between the RIS and the user, is estimated on two distinct timescales (i.e., slow and fast time-varying timescales), respectively. Naturally, the effectiveness of this scheme is owing to the slow time-varying channel caused by the known and fixed location of the BS and the RIS. Furthermore, {\bf{Characteristic 1}} indicates that for all users, the common channel only has to be estimated once. Upon completing the channel estimation for BSs-RISs, the channel estimation for RISs-Uers can be accomplished by multi-BS cooperation based on {\bf{Characteristic 2}}.

Considering the above, the protocol and frame structure of two-timescale channel estimation for RIS-assisted cell-free systems need to be redesigned on the basis of the TS mode. To put it crudely, for each RIS, the common channels $\mathbf{F}$ of all users are estimated in the first timescale, and then the individual channel of each user is estimated by the cooperation of multiple BSs in $\bar{Q^\prime}$ sub-frames, where each sub-frame consists of $T$ symbol durations, $T\geq K$. 

There has been several research \cite{SVDCE,two-timescale} on how to obtain the channel from the BS to the RIS, and they can be extended to the case of multiple BSs and RISs\footnote{Improved techniques are needed in the future for the multi-BS multi-RIS situation.}. 
In this work, we assume that channels between BSs and RISs, i.e., $\mathbf{F}_{mn}$, are known to achieve the implement of {\bf{characteristic 2}}. This shows that we are primarily concerned with the second timescale channel estimation.

Considering the TS mode without taking into account RIS's subscript, in line with Eqn. (\ref{Y}), the received signal at the $m$-th BS sent by the $k$-th user in $\bar{Q^\prime}$ sub-frames can be denoted by
\begin{equation}\label{Y2}
	\widetilde{\mathbf{Y}}_{m,k}
	= \begin{bmatrix} {{{\mathbf{F}}_{m}^{H}}\mathbf{V}^1} \\
		\ {{{\mathbf{F}}_{m}^{H}}\mathbf{V}^2}\\
		\vdots\\
	 {{{\mathbf{F}}_{m}^{H}}\mathbf{V}^{\bar{Q^\prime}}}
	\end{bmatrix}{{\mathbf{h}}_{k}} +
	\begin{bmatrix}\widetilde{\mathbf{W}}_{m,1}\\
		\widetilde{\mathbf{W}}_{m,2}\\
		\vdots\\
		\widetilde{\mathbf{W}}_{m,\bar{Q^\prime}}
	\end{bmatrix}.
\end{equation}
 Using the VCR expression of $\mathbf{h}_k=\widetilde{\mathbf{A}}_{\rm R}\mathbf{x}_k$, we yield
\begin{equation}\label{CS2}
	\widetilde{\mathbf{Y}}_{m,k}=\mathbf{\Phi}_m\widetilde{\mathbf{A}}_{\rm R}\mathbf{x}_k+\widetilde{\mathbf{W}}_m,
\end{equation}
where $\mathbf{\Phi}_m=[(\mathbf{F}_m^H\mathbf{V}^1)^T,\cdots,(\mathbf{F}_m^H\mathbf{V}^{\bar{Q^\prime}})^T]^T\in\mathbb{C}^{{\bar{Q^\prime}} J_m\times L}$, $\widetilde{\mathbf{A}}_{\rm R}\in\mathbb{C}^{L\times G_r}$ is the dictionary that can be expressed as Eqn. (\ref{D_A}). The nonzero entries of $\mathbf{x}_k\in\mathbb{C}^{G_r\times 1}$ correspond the entries of AoAs and gains of the $k$-th user's channel, and $\widetilde{\mathbf{W}}_m=[\widetilde{\mathbf{W}}_{m,1}^T,\cdots, \widetilde{\mathbf{W}}_{m,{\bar{Q^\prime}}}^T]^T\in\mathbb{C}^{{\bar{Q^\prime}} J_m\times 1}$. This issue can be esaily addressed by each BS's individual estimation.

On the other hand, according to {\bf{characteristic 2}}, the signals received by all BSs can be used to jointly estimate $\mathbf{x}_k$. As the number of BSs rises, so does the number of measurements, which leads to greater estimation performance. By stacking all BSs' signals, we yield
\begin{equation}\label{CS3}
	\begin{aligned}
		&\underset{\mathbf{x}_{k}}{\rm arg \ min} \ \Vert \mathbf{x}_{k} \Vert_0 \\
	&	 {\rm s.t. \ }\Vert\widetilde{\mathbf{Y}}_{k}-\widetilde{\mathbf{\Phi}}\mathbf{x}
		\Vert_2<\epsilon,
	\end{aligned}
\end{equation}
where $\widetilde{\mathbf{Y}}_{k}=[\widetilde{\mathbf{Y}}_{1,k}^T,\cdots,\widetilde{\mathbf{Y}}_{N_B,k}^T]^T\in\mathbf{C}^{{\bar{Q^\prime}} \mathcal{J}\times L}$, $\mathcal{J}=\sum_{m=1}^{N_B}J_m$ is the total number of antennas for all BSs and $\widetilde{\mathbf{\Phi}}=[(\mathbf{\Phi}_1\widetilde{\mathbf{A}}_{\rm R})^T,\cdots,(\mathbf{\Phi}_{N_B}\widetilde{\mathbf{A}}_{\rm R})^T]^T\in\mathbb{C}^{{\bar{Q^\prime}} \mathcal{J}\times G_r}$ is the multi-BS cooperative sensing matrix.
The increment of measurements reveals that multi-BS cooperation may accomplish channel estimation with little pilot overhead.
\section{Simulation Results}
This section conducts several experiments to demonstrate the efficacy of our proposed cascaded and two-timescale channel estimation methodologies. 
 We set $N_B=3$, $N_R=3$, $K=8$, $J_1=\cdots=J_{N_B}=16$ and $L_1=\cdots=L_{N_R}=128$. The codebook size $G_t$ and $G_r$ are set to 512. In addition, we assume the locations of all BSs, RISs and uers are placed randomly. $\phi_{m,n,p}$, $\varphi_{n,k,b}$ and $\vartheta_{m,n,p}$ are uniformly distributed on $[0,2\pi)$, $P_{f,m,n}=P_{h,n,k}=3$, $\beta_{m,n,p}$ and $\gamma_{n,k,b}$ follow complex Gaussian distribution with unit power. The signal to noise (SNR) is defined as SNR $=\sigma_p^2/\sigma_n^2$. Besides, the RIS phase matrix follows complex Gaussian distribution with unit power.

\subsection{Cascaded Channel Estimation}
The normalized mean square error (NMSE) of $\mathbf{G}$ is defined as follows:
\begin{equation}
	{\rm NMSE}_{\mathbf{G}}=\frac{1}{N_BN_RK}\sum_{m=1}^{N_B}\sum_{n=1}^{N_R}\sum_{k=1}^{K}\frac{\Vert\widehat{\mathbf{G}}_{mnk}-\mathbf{G}_{mnk}\Vert_F^2}{\Vert \mathbf{G}_{mnk}\Vert_F^2}.
\end{equation}
\begin{figure}[htbp]
	\centering
	\includegraphics[width=7cm,height=5.3cm]{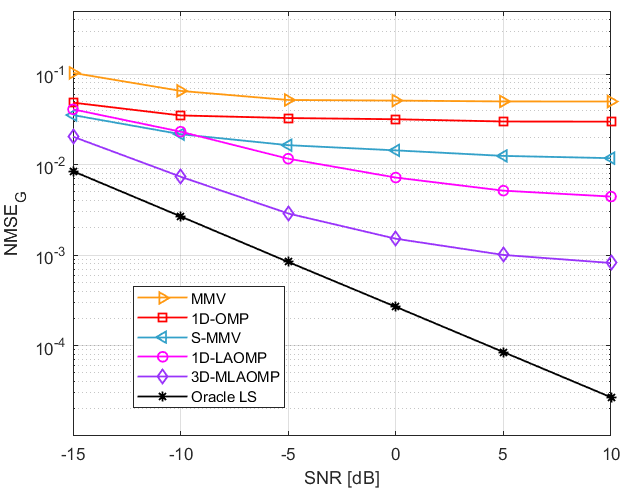}
	\centering
	\caption{NMSE of different methods versus SNR with $\bar{Q}=32$.  }
	\label{cascaded}
\end{figure}

We employ the following benchmark schemes as comparisons to our proposed scheme (3D-MLAOMP), where the look ahead parameter $U=3$ is set for AoD estimation and $U=9$ for AoA estimation. 1) 1-D CS schemes without considering the characteristic of multiuser cascaded channel, i.e., OMP \cite{CS1} and LAOMP with $U=9$; 2) MMV and subspace estimation-based MMV (S-MMV) \cite{CS2}\footnote{This research futher considered a two-step subspace based multi-user joint estimation (S-MJCE) method which resulted in an extraordinarily high level of computational complexity. As a result, it is omitted from consideration in RIS-assisted cell-free systems.}; 3) Oracle LS method with true AoAs/AoDs, which serves as the lower bound and cannot be exceeded.
\figurename{\ref{cascaded}} exhibits the NMSE comparison of different schemes versus SNR, where the number of measurements $\bar{Q}=32$ is set for all schemes. This figure demonstrates that the critical nature of multiuser channel is beneficial for multi-user cascaded channel estimation.
\subsection{Two-Timescale Channel Estimation}
Since we focus on the second timescale channel estimation, the NMSE of $\mathbf{h}$ is defined as follows:
\begin{equation}
	{\rm NMSE}_{\mathbf{h}}=\frac{1}{N_RK}\sum_{n=1}^{N_R}\sum_{k=1}^{K}\frac{\Vert\widehat{\mathbf{h}}_{nk}-\mathbf{h}_{nk}\Vert_2^2}{\Vert \mathbf{h}_{nk}\Vert_2^2}.
\end{equation}
\begin{figure}[htbp]
	\centering
	\includegraphics[width=7cm,height=5.3cm]{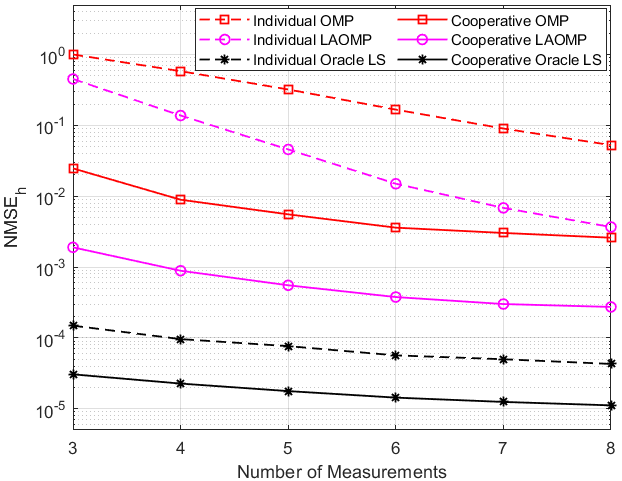}
	\centering
	\caption{NMSE$_\mathbf{h}$ of different individual and cooperative schemes versus the number of measurements.}
	\label{twotimescale}
\end{figure}

To highlight the strength of our proposed multi-BS cooperative estimation, we take individuality-based estimation methods as the benchmark, where the performance of individuality-based estimation is the average of all BSs' NMSE$_\mathbf{h}$. 
As shown in \figurename{\ref{twotimescale}}, where SNR $=10$ dB and $U=9$ are set, we can see that cooperation-oriented schemes outperform strategies based on individuality. This demonstrates the superiority of our proposed multi-BS cooperative estimation strategy.

\section{Conclusions}
By focusing on the multi-BS multi-user scenario in RIS-assisted cell-free systems, we investigate the feasibility of multi-BS cooperation and joint multi-user estimation. As for cascaded channel estimation, we suggest a 3D-MMV framework to jointly estimate cascaded AoDs for all users in light of {\bf{characteristic 1}}. We further utilize tensor contraction to present a 3D-MLAOMP algorithm.
Besides, it seems as if {\bf{characteristic 2}} is tough to implement in cascaded channel estimation. Therefore, we take into account two-timescale channel estimation with the usage of {\bf{characteristic 1}} and {\bf{2}}, followed by the development of a pilot-reduced CS framework. Finally, simulation results demonstrate the effectiveness of our proposed schemes. In future work, we will investigate how to efficiently derive the first timescale channel estimation via multi-BS cooperation. Additionally, non-orthogonal pilot sequences-based RIS-assisted cell-free channel estimation in high mobility scenarios is worth investigating.

\ifCLASSOPTIONcaptionsoff
  \newpage
\fi

\bibliographystyle{IEEEtran}

\end{document}